\documentclass[prd,amsmath,amssymb,showpacs,superscriptaddress,nofootinbib,twocolumn]{revtex4-1}
\usepackage{graphicx}
\usepackage{epsfig,graphics,subfigure,psfrag,amsmath,amssymb}
\usepackage{lineno}
\usepackage{dcolumn}
\usepackage{bm}
\usepackage{overpic}
\usepackage{xspace}

\usepackage{rotating}
\usepackage{color}
\usepackage{multirow}
\usepackage{colortbl}
\usepackage{epstopdf}
\usepackage[colorlinks,linkcolor=blue,anchorcolor=blue,citecolor=blue]{hyperref}

\newcommand{\jpsi}{J/\psi}

\newcommand{\kk}{K^{+}K^{-}}
\newcommand{\ks}{K_{S}^{0}}

\newcommand{\ppphi}{p\bar p\phi}
\newcommand{\psip}{\psi(3686)}
\newcommand{\ppbar}{p\bar p}
\newcommand{\pphi}{p\phi}
\newcommand{\pbarphi}{\bar p\phi}

\newcommand{\BESIII}{BES\uppercase\expandafter{\romannumeral3}\xspace}

\begin{document}
\title{Observation of $\psi(3686)\to p\bar{p}\phi$}
\author{M.~Ablikim$^{1}$, M.~N.~Achasov$^{10,d}$, P.~Adlarson$^{59}$, S. ~Ahmed$^{15}$, M.~Albrecht$^{4}$, M.~Alekseev$^{58A,58C}$, A.~Amoroso$^{58A,58C}$, F.~F.~An$^{1}$, Q.~An$^{55,43}$, Y.~Bai$^{42}$
% ...
\\
      \vspace{0.2cm}
      (BESIII Collaboration)\\
      \vspace{0.2cm} {\it
$^{1}$ Institute of High Energy Physics, Beijing 100049, People's Republic of China\\
% ...
$^{4}$ Bochum Ruhr-University, D-44780 Bochum, Germany\\
% ...
$^{10}$ G.I. Budker Institute of Nuclear Physics SB RAS (BINP), Novosibirsk 630090, Russia\\
% ...
$^{15}$ Helmholtz Institute Mainz, Johann-Joachim-Becher-Weg 45, D-55099 Mainz, Germany\\
% ...
$^{42}$ Southeast University, Nanjing 211100, People's Republic of China\\
$^{43}$ State Key Laboratory of Particle Detection and Electronics, Beijing 100049, Hefei 230026, People's Republic of China\\
% ...
$^{55}$ University of Science and Technology of China, Hefei 230026, People's Republic of China\\
% ...
$^{58}$ (A)University of Turin, I-10125, Turin, Italy; (B)University of Eastern Piedmont, I-15121, Alessandria, Italy; (C)INFN, I-10125, Turin, Italy\\
$^{59}$ Uppsala University, Box 516, SE-75120 Uppsala, Sweden\\
% ...
$^{d}$ Also at the Novosibirsk State University, Novosibirsk, 630090, Russia\\
% ...
}
}

\date{\today}

\begin{abstract}
  Using a data sample of $4.48\times10^{8}$ $\psip$ events collected with the BESIII detector, we present a first observation of $\psi(3686)\to p\bar{p}\phi$,  and we measure its branching fraction to be $[6.06\pm0.38 ($stat.$) \pm 0.48 ($syst.$)]\times10^{-6}$.
  In contrast to the earlier discovery of a threshold enhancement in the $p\bar{p}$-mass spectrum of the channel $J/\psi\to\gamma p\bar p$, denoted as $X(p\bar{p})$, we do not find a similar enhancement in $\psi(3686)\to p\bar{p}\phi$. An upper limit of $1.82\times10^{-7}$ at the $90\%$ confidence level on the branching fraction of $\psi(3686)\to X(p\bar{p})\phi\to p\bar{p}\phi$ is obtained.

\end{abstract}

\pacs{13.66.Bc, 14.40.Be}
\maketitle

\section{INTRODUCTION}

An intriguing enhancement near the $\ppbar$-mass threshold, referred to as the
$X(\ppbar)$, was discovered by BES in the channel $J/\psi\to\gamma p \bar{p}$~\cite{cite01}
and subsequently confirmed by CLEO~\cite{cite02}
and BESIII~\cite{cite03}. A more recent partial-wave amplitude analysis of $J/\psi\to\gamma\ppbar$~\cite{cite04}
supports the existence of the structure and concludes to a spin-parity assignment of $J^{PC} = 0^{-+}$.
There is no experimental evidence of such an enhancement in radiative $\Upsilon(1S)\to\gamma\ppbar$~\cite{cite05}
decay nor in the $J/\psi\to\omega\ppbar$ decay~\cite{cite06}.
It is tempting to associate this enhancement with
the $X(1835)$, a resonance that was recently confirmed
by BESIII~\cite{cite09} after it was first observed
in $J/\psi\to\gamma\pi^{+}\pi^{-}\eta'$ decay~\cite{cite10}.
Whether or not the $\ppbar$-mass threshold enhancement and the $X(1835)$
are related to the same source still needs further study.
As a result, lots of theoretical speculations have been proposed
to interpret the nature of this structure, including
the quasibound nuclear baryonium~\cite{cite11,cite12},
a multiquark resonance~\cite{cite13} or an effect caused by
final-state interaction (FSI)~\cite{cite14,cite15} near the proton-antiproton production threshold.

Most recently, BESIII reported the study of $J/\psi\to\ppphi$~\cite{cite20} and no evidence of a
near-threshold enhancement in the $\ppbar$-mass spectrum was found. Moreover,
no significant signatures of resonances in the $\pphi$ or $\pbarphi$ mass spectra were observed.
 For the decay of $\psip\to\ppphi$, BES reported an upper limit on the branching fraction $\mathcal{B}(\psip\to\ppphi)$
 of $2.6\times10^{-5}$ at the $90\%$ confidence level (C.L.)~\cite{citeBESI}.
 The latest measurement came from \mbox{CLEO~\cite{cite21}},
 who reported an upper limit on the branching fraction $\mathcal{B}(\psip\to\ppphi)$ of $2.4\times10^{-5}$
 at the $90\%$ C.L..
 These experimental observations, together with similar results found in different decays,
 give rise to a discussion on the nature of the threshold effect and stimulate theoretical developments.

 In this work, we report on the data
 analysis of the charmonium decay $\psip\to\ppphi$.
 The data have been obtained with the BESIII detector at the BEPCII storage ring at which a total
 of $(4.481\pm0.029)\times10^{8}$ $\psip$ events~\cite{cite22} were produced in electron-positron annihilations.
 The aim of this work is to search for a near-threshold enhancement in the
 $\ppbar$-mass spectrum and to search for $\pphi(\pbarphi)$  resonances that might hint to the existence
 of pentaquarks with hidden strangeness. Moreover,
 we measured the branching fraction of the process $\psip\to\ppphi$
 which allows us to inspect the `$12\%$ rule' proposed in 1975~\cite{cite23}.
 The rule is based on perturbative quantum chromodynamics (QCD) calculations,
 in which the ratio of the branching fractions of $\psi(3686)$ and $J/\psi$ into the same final hadronic state is given by

   \begin{equation}
   \begin{small}
   Q = \frac{B_{\psip\to h}}{B_{J/\psi\to h}} = \frac{B_{\psip\to l^{+}l^{-}}}{B_{J/\psi\to l^{+}l^{-}}} =  (12.4\pm0.4)\%.
   \end{small}
 \end{equation}

\section{DETECTOR AND MONTE CARLO SIMULATION}
The BESIII detector is a magnetic
spectrometer~\cite{Ablikim:2009aa} located at the Beijing Electron
Positron Collider (BEPCII)~\cite{Yu:IPAC2016-TUYA01}. The
cylindrical core of the BESIII detector consists of a helium-based
 multilayer drift chamber (MDC), a plastic scintillator time-of-flight
system (TOF), and a CsI(Tl) electromagnetic calorimeter (EMC),
which are all enclosed in a superconducting solenoidal magnet
providing a 1.0~T
magnetic field. The solenoid is supported by an
octagonal flux-return yoke with resistive plate counter muon
identifier modules interleaved with steel. The acceptance of
charged particles and photons is 93\% over $4\pi$ solid angle. The
charged-particle momentum resolution at $1~{\rm GeV}/c$ is
$0.5\%$, and the $dE/dx$ resolution is $6\%$ for the electrons
from Bhabha scattering. The EMC measures photon energies with a
resolution of $2.5\%$ ($5\%$) at $1$~GeV in the barrel (end cap)
region. The time resolution of the TOF barrel part is 68~ps, while
that of the end cap part is 110~ps.

Simulated samples produced with the {\sc
geant4}-based~\cite{geant4} Monte Carlo (MC) package which
includes the geometric description of the BESIII detector and the
detector response, are used to determine the detection efficiency
and to estimate the backgrounds. The simulation includes the beam
energy spread and the initial-state radiation (ISR) in the $e^+e^-$
annihilations modelled with the generator {\sc
kkmc}~\cite{ref:kkmc}. The inclusive MC sample consists of the production of the $\psi(3686)$
resonance, and the continuum processes incorporated in {\sc kkmc}.
The known decay modes are modelled with {\sc
evtgen}~\cite{ref:evtgen} using branching fractions taken from the
Particle Data Group (PDG)~\cite{pdg}, and the remaining unknown decays
from the charmonium states with {\sc
lundcharm}~\cite{ref:lundcharm}. The final-state radiation (FSR)
from charged final-state particles is incorporated with the {\sc
photos} package~\cite{photos}.
The background is studied
using a sample of $5.06\times10^{8}$ inclusive $\psip$ MC events.
 The analysis is performed in the framework of the BESIII offline software system
(BOSS)~\cite{ref:boss} incorporating
the detector calibration, event reconstruction and data storage.

\section{DATA ANALYSIS}
\subsection{Event selection and background analysis}

The $\psip\to\ppphi$ reaction is identified with the
$\phi$ subsequently decaying into $\kk$ resulting in a
 final state of four charged tracks, namely $\ppbar\kk$.
 The charged tracks must have been detected in the
 active region of the MDC, corresponding to $|\cos\theta| < 0.93$,
  where $\theta$ is the polar angle of the charged track
  with respect to the beam direction. Moreover, the
  tracks are required to pass within $\pm$10 cm of the interaction
  point in the beam direction and within $\pm$1 cm
   in the plane perpendicular to the beam. Two
   of the charged tracks are identified as a proton and an antiproton by using combined TOF and $dE/dx$
   information. To improve the detection efficiency, the
   events with at least one $K^{+}(K^{-})$ are selected for further analysis. Thus, the candidate events are
   required to have three or four charged tracks. A one-constraint (1C) kinematic fit is subsequently performed under the
   hypothesis of  $\psip\to\ppbar\kk$ , where $K^{+}$ or $K^{-}$
   is treated as a missing particle with the nominal mass of a kaon. For the events with both kaons
   detected, two 1C kinematic fits are performed assuming a missing $K^{+}$ or $K^{-}$.
   The one with the least $\chi^{2}_{1C}$ is retained.
   To suppress background events, the $\chi^{2}_{1C}$ is required to be less than 10.

\begin{figure}[htbp]
  \centering
  \vskip -0.2cm
  \hskip -0.4cm \mbox{
  \begin{overpic}[width=0.45\textwidth]{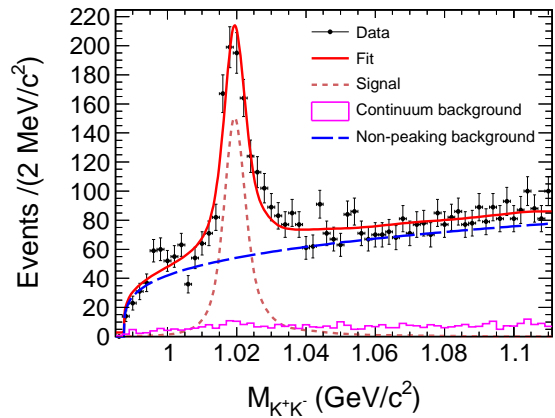}
  \end{overpic}
  }
  \vskip -0.3cm
  \hskip 0.5cm
  \caption{\label{Mkk}
  Fit to $\kk$ invariant-mass spectrum. The
  dots with error bars represent the data, the red solid
line is the global fit result, the brown short dashed line represent the signal shape,
the pink histogram is the contribution of the continuum background,
and the blue long dashed line reflects the non-peaking background.
      }
\end{figure}

\begin{figure}[htbp]
  \centering
  \vskip -0.2cm
  \hskip -0.4cm \mbox{
  \begin{overpic}[width=0.45\textwidth]{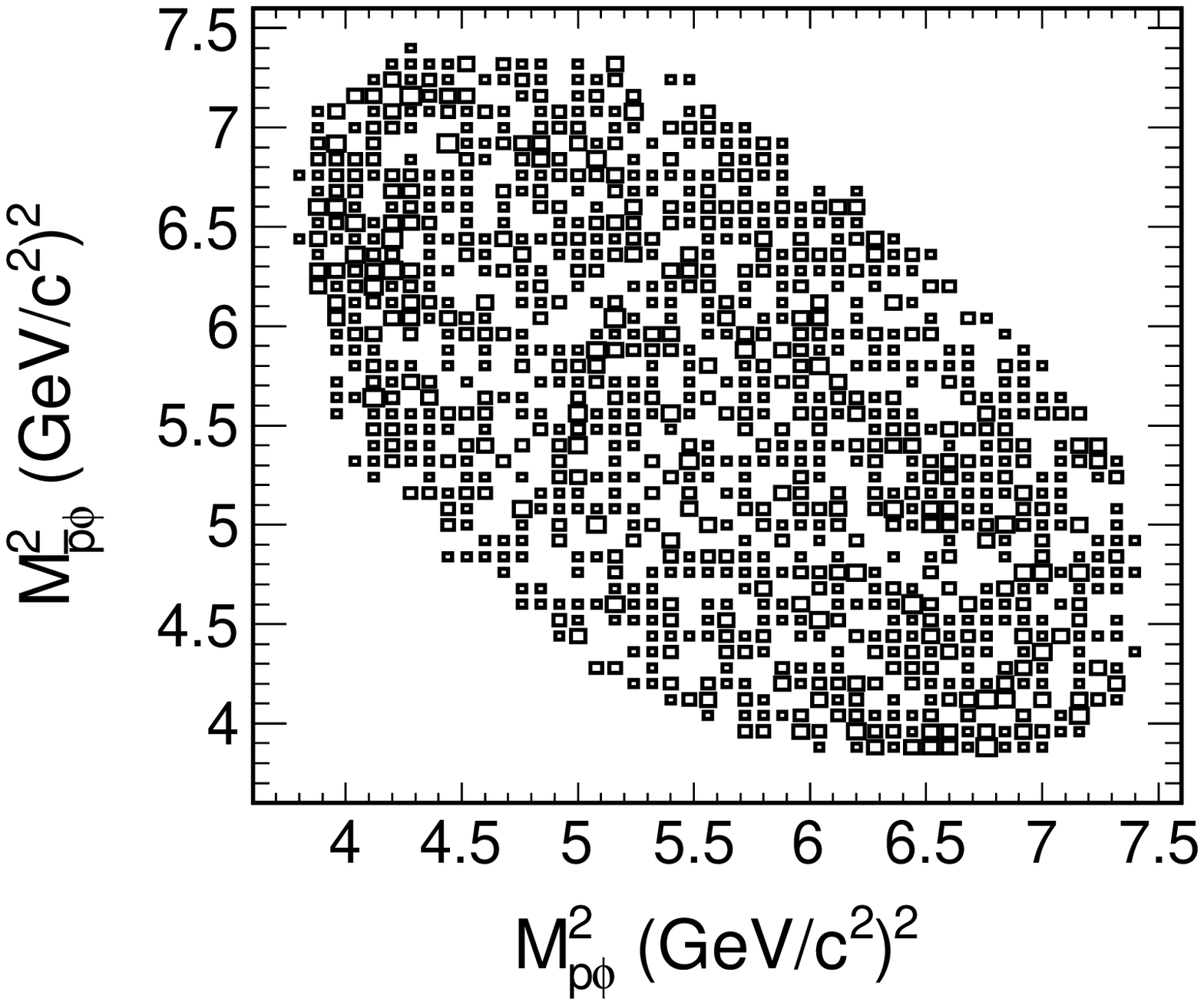}
  \end{overpic}
  }
  \vskip -0.3cm
  \hskip 0.5cm
  \caption{\label{dalitz}
  Dalitz plot for $\psip\to\ppphi$
   for the events with a $\kk$
   invariant mass that falls within the $\phi$-mass region
   (1.005 GeV/$c^{2} < M_{\kk} <$ 1.035 GeV/$c^{2}$).
      }
\end{figure}

\begin{figure*}[htbp]
  \centering
  \vskip -0.2cm
  \hskip -0.4cm \mbox{
  \begin{overpic}[width=0.45\textwidth]{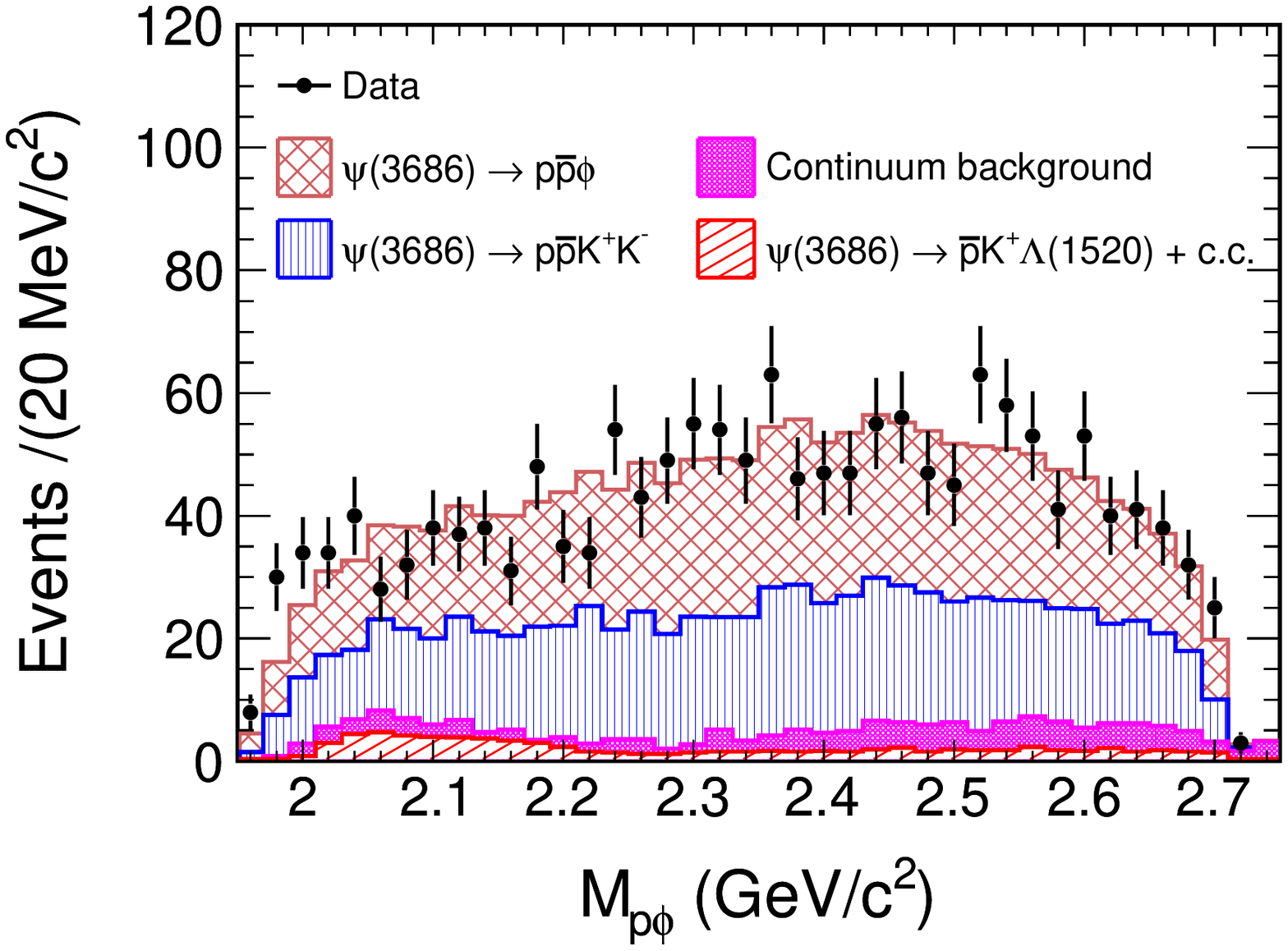}
  \put(85,59){{(a)  }}
  \end{overpic}
  \begin{overpic}[width=0.45\textwidth]{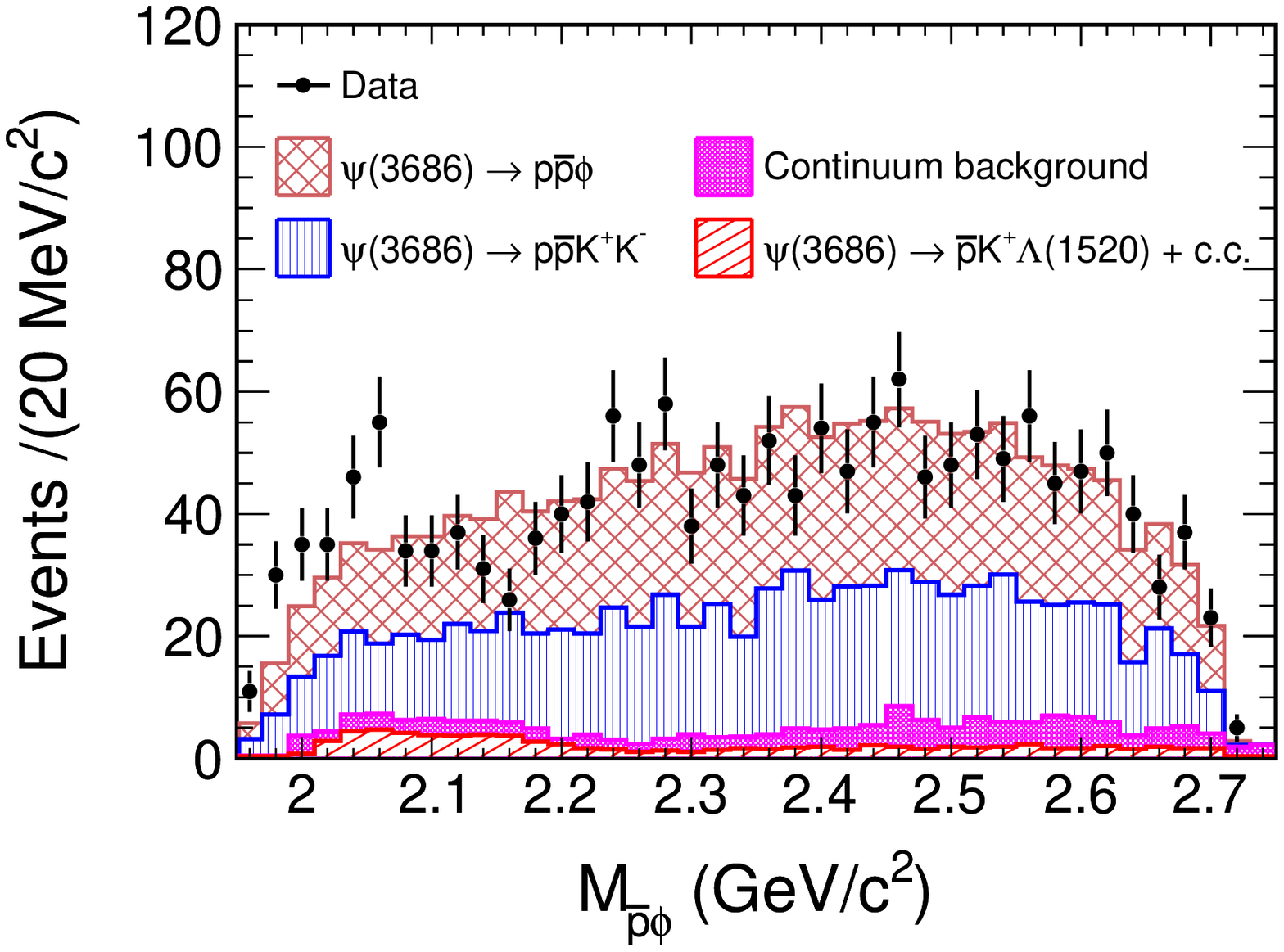}
  \put(85,59){{(b)  }}
  \end{overpic}
  }
  \vskip -0.3cm
  \hskip 0.5cm
  \caption{\label{Mpphi}
 The invariant-mass distribution of (a) $p\phi$ and (b) $\bar p\phi$.
 The dots with error bars denote the data; the contributions for each component are
 indicated as the hatched histograms.
      }
\end{figure*}

   The potential backgrounds are investigated using
   the inclusive $\psip$ MC sample. Besides the irreducible
   backgrounds from the non-resonant decay $\psip\to\ppbar\kk$,
   the reducible backgrounds are dominated by the processes involving
   $\Lambda (\bar \Lambda)$ intermediate states.
   To suppress the above backgrounds, all other charged tracks
   except for the selected proton, antiproton and kaon candidates
   are assumed to be pions, and events are excluded if any combination of
   $p\pi^{-}$ or $\bar p\pi^{+}$  has an invariant mass lying in the range
   $|M_{p\pi^{-} (\bar p\pi^{+})} - M_{\Lambda (\bar\Lambda)}|< 3~$MeV/$c^{2}$.
   There are also some background events found originating from the process $\psi(3686)\to
   \bar{p} K^{+}\Lambda(1520) + c.c.$ with $\Lambda(1520)\to pK$.
   A MC sample is generated to describe its shape, and the number of background events of  $\psi(3686)\to
   \bar{p} K^{+}\Lambda(1520)$ is expected to be $40\pm21$,
   which is estimated by a fit to the measured $pK^{-}$ invariant-mass spectrum.
   The signal shape of the $\Lambda(1520)\to pK$ is modeled with a Breit Wigner (BW) function,
   and the background is described with a second-order Chebychev polynomial function.
   Only the background from the continuum process $e^{+}e^{-}\to\ppphi$
   was found to have a peaking structure underneath the $\phi$-signal region.
   This contribution from this background is studied using
   the off-resonance samples taken at $\sqrt{s} = 3.773$ GeV,
   and its absolute magnitude is determined
   according to the formula
   $N =N^{survive}_{3773}\cdot\frac{\mathcal{L}_{\psi(3686)}}{\mathcal{L}_{3773}}\cdot\frac{\sigma_{\psi(3686)}}{\sigma_{3773}}\cdot \frac{\varepsilon_{\psi(3686)}}{\varepsilon_{3773}}$,
   where $N^{survive}_{3773}$ is the number of events which remained in the off-resonance samples
    after applying the same event selections that are used to identify $\psi(3686)\to\ppphi$.
    $\mathcal{L}$, $\sigma$ and $\varepsilon$ refer to the integrated luminosities
    ($\mathcal{L}_{\psi(3686)} = 668.55$~pb$^{-1}$~\cite{cite22},
        $\mathcal{L}_{3773} = 2931.8$~pb$^{-1}$~\cite{N3773},
    the cross sections and the detection efficiencies of the data samples taken at the two corresponding center-of-mass energies, respectively.
   Figure~\ref{Mkk} shows the $\kk$ invariant-mass spectrum after applying all the
   selection criteria mentioned above.
   Note that a clear signal corresponding to the decay $\phi\to\kk$ is visible in the spectrum.
    Figure~\ref{dalitz} shows the Dalitz plot of $\psip\to\ppphi$ for the events with a $\kk$
   invariant-mass that falls within the $\phi$-mass region
   (1.005~GeV/$c^{2} < M_{\kk} <$ 1.035~GeV/$c^{2}$). The data show no evident resonance structures.
   Figure~\ref{Mpphi} shows its projections on the $\pphi$ and $\pbarphi$
   invariant-mass distributions.
   These distributions show that the data are well described by
   a phase-space distribution of the signal channel together with the continuum background and non-peaking background.

\subsection{Measurement of $\mathcal{B}(\psip\to\ppphi)$}

 The $\phi$-signal yields are obtained from an extended
 unbinned maximum-likelihood fit to the $\kk$
 invariant-mass spectrum in the range of $[0.985, 1.115]$~GeV/$c^{2}$.
 In the fit, the $\phi$ signal component is modeled by the MC-simulated signal shape
 convoluted with a Gaussian function to account for
 the difference in the mass resolution between
 data and MC simulation. The MC sample, $\psi(3686)\to p\bar{p}\phi$,
 is generated according to a phase-space assumption.
 The background contribution from the continuum process $e^{+}e^{-}\to\ppphi$ is
   obtained as discussed above and
   its shape and yield are fixed in the fit.
 The other background events
 are parameterized by a modified ARGUS function~\cite{cite32}.
 The parameters of the Gaussian function and the ARGUS
  function are left free in the fit. The fit, shown
  in Fig.~\ref{Mkk}, yields $N_{obs} = 753\pm47$ signal events.
  The statistical significance is found to be 21~$\sigma$,
which is determined from the change in $-2\ln L$ in the fits of mass spectrum with and without
   assuming the presence of a signal while considering the change in degrees of freedom of the fits.

  The branching fraction of $\psip\to\ppphi$ is calculated with,
   \begin{equation}
    \begin{split}
   \mathcal{B}(\psip&\to\ppphi)\\
   &=\frac{N_{obs}}{N_{\psip}\times \mathcal{B}(\phi\to\kk)\times\varepsilon},
   \end{split}
 \end{equation}

  \noindent where $N_{obs}$ is the number of the observed signal events which comes from the fit.
  $N_{\psip}$ is the total number of $\psip$ events.
  The branching fraction of $\phi\to\kk$, $\mathcal{B}(\phi\to\kk) = (49.2\pm0.5)\%$,
  is taken from the PDG~\cite{pdg}. $\varepsilon$ is the detection efficiency.
  To obtain a reliable detection efficiency, the MC sample of $\psip\to\ppphi$,
  distributed according to a phase-space assumption,
is weighted to match the distribution of the background-subtracted data with the mass distribution of
$p\bar{p}$, and the average detection efficiency is determined to be 56.4\%.
  The branching fraction, $\mathcal{B}(\psip\to\ppphi)$, is measured to be $(6.06\pm0.38\pm0.48)\times10^{-6}$,
  where the uncertainties are the statistical and systematic uncertainty, respectively.
  The systematic uncertainties will be discussed in detail in the following section.

\subsection{Systematic uncertainties}

The systematic uncertainties that affect the branching-fraction measurement can be divided into two categories.
 The first category is given by the uncertainties in the track reconstruction,
 the particle identification (PID), 1C kinematic fit,  and $\Lambda/\bar{\Lambda}$ veto efficiency.
 The other category comprises the uncertainties which originate from the fit of the mass spectrum,
 the weighting procedure,
 the cited branching fraction of the decay of the intermediate state,  and the total number of $\psi(3686)$ events.

 The difference in the efficiencies of the track reconstruction for $p/\bar{p}$ between MC and data
 is studied using a clean sample of $J/\psi\to p\bar{p}\pi^{+}\pi^{-}$ and found to be less than
1.0\% per track. For the $K^{\pm}$, the systematic uncertainty is
studied using a clean control sample of $\jpsi\to\ks K^{\pm}\pi^{\mp}$.
1.0\% per tracking is taken as the systematic uncertainty for the tracking efficiency~\cite{ywccite33}.

The PID efficiency of $p/\bar{p}$ is also studied from the same data sample of $J/\psi\to p\bar{p}\pi^{+}\pi^{-}$.
The results indicate that the $p/\bar{p}$ PID efficiency for data agrees with the MC simulation within 1\%.
The PID efficiency for the kaon is measured in the clean channel $J/\psi\rightarrow K^+ K^-\eta$.
It is found that the difference between the PID efficiency of data and MC is less than 1\% for each kaon.
In this analysis, three charged tracks are required to be identified as a proton,
an anti-proton and a kaon. Hence, 3\% is taken as the systematic uncertainty associated with the PID.

With a clean control sample of $\psi(3686) \to pK^-\bar{\Lambda} + c.c.$,
the systematic uncertainty of the 1C kinematic fit is estimated to be 3.4\%
by calculating the difference of ratio of signal yields with $\chi^2_{1C}$ cut
and without 1C kinematic fit between MC simulation and data.

To veto the $\Lambda/\bar{\Lambda}$ background events,
$|M_{p\pi^-/\bar{p}\pi^+} - M_{\Lambda/\bar{\Lambda}}| > 3 $ MeV/${c^2}$
is required. An alternative choice of
$|M_{p\pi^-/\bar{p}\pi^+} - M_{\Lambda/\bar{\Lambda}}| > 10$  MeV/${c^2}$
is used to remeasure the branching fraction. A difference of 1.1\%
is found and taken as the corresponding systematic uncertainty.

The $\phi$-signal yields are obtained by fitting the $K^{+}K^{-}$ invariant-mass spectrum.
Systematic uncertainties related to the fit have been estimated by using different signal
and background shapes, alternative fit ranges,
and by taking into consideration an additional resonant structure.
To estimate the uncertainty from the modeling of the $\phi$-signal shape,
an alternative fit with an acceptance-corrected BW function to describe the $\phi$-signal has been performed.
To estimate the uncertainty due to the background shape,
a function of $f(M) = (M - M_a)^c(M_b - M)^d$
is used instead of the modified ARGUS function, where,
$M_a$ and $M_b$ are the lower and upper edges of the mass distribution, respectively, and
$c$ and $d$ are free parameters. In the $K^+K^-$ invariant-mass distribution,
we observed a small bump around 1~GeV/$c^2$.
Although this structure might be due to statistical fluctuations,
we considered the possibility of an additional resonance.
We, therefore, fitted the distribution with an extra BW function convolved with a Gaussian function.
The change of signal yield in the different fit is taken as the corresponding systematic uncertainty.
The quadratic sum of the four individual uncertainties is taken as the systematic uncertainty
related with the mass spectrum fit, and it is found to be 5.5\%.

To obtain a reliable detection
efficiency, the MC sample modeled using a phase-space distribution is weighted to
match the distribution of the background-subtracted data.
To consider the effect on the statistical
fluctuations of the signal yield in the data, a set of toy-MC samples are used to estimate the
detection efficiencies. With the reweighting, a maximum deviation in detection efficiencies of
1.0\% is found and quoted as the corresponding systematic uncertainty.

The branching fraction uncertainty of the intermediate decay $\phi \to K^+K^-$, 1.0\%,
is taken from the PDG and the uncertainty of the number of $\psi(3686)$ events is 0.6\%~\cite{cite22}.

In Table~\ref{sys}, a summary is shown of all contributions to the systematic uncertainties on the
branching fraction measurements.
The total systematic uncertainty is given by the quadratic sum of the individual contributions,
assuming all sources to be independent.

\begin{table}[!htbp]
  \centering
  \caption{\label{sys}
  Sources of relative systematic uncertainties and their contributions to the branching fractions
 and upper limits (in \%).}
\begin{tabular}{ l c c }
  \hline
  \hline
  Sources                       & $p\bar{p}\phi$ & ~~~$X(p\bar{p})\phi$ \\ \hline
  MDC tracking                  & 3.0 & 3.0 \\
  PID efficiency                & 3.0 & 3.0 \\
  1C kinematic fit              & 3.4 & 3.4 \\
  $\Lambda(\bar{\Lambda})$ veto & 1.1 & 1.1 \\
  Mass spectrum fit             & 5.5 & --- \\
  Weighting procedure           & 1.0 & --- \\
  $\mathcal{B}(\phi \to K^+K^-)$          & 1.0 & 1.0 \\
  Number of $\phi(3686)$ events & 0.6 & 0.6 \\\hline
  Total                         & 8.0 & 5.7 \\
  \hline
  \hline
\end{tabular}
\end{table}

\subsection{Upper limit of $p\bar{p}$ mass threshold enhancement}

\begin{figure}[htbp]
  \centering
  \vskip -0.2cm
  \hskip -0.4cm \mbox{
  \begin{overpic}[width=0.45\textwidth]{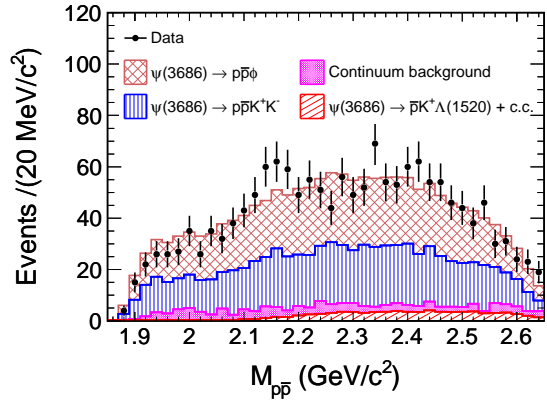}
  \end{overpic}
  }
  \vskip -0.3cm
  \hskip 0.5cm
  \caption{\label{Mppbar}
  The $\ppbar$ invariant-mass distribution. The dots with error bars denote the data; the contributions for each component are
 displayed as the hatched histograms.
      }
\end{figure}

\begin{figure}[htbp]
  \centering
  \vskip -0.2cm
  \hskip -0.4cm \mbox{
  \begin{overpic}[width=0.45\textwidth]{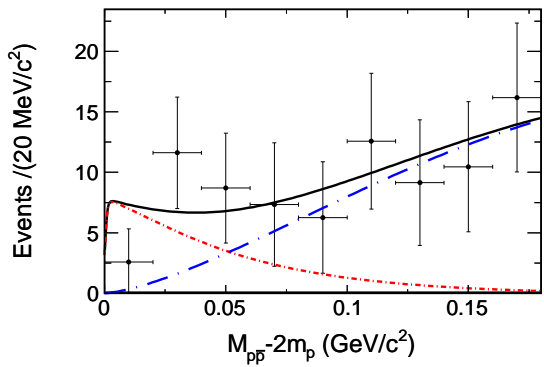}
  \end{overpic}
  }
  \vskip -0.3cm
  \hskip 0.5cm
  \caption{\label{Fit}
   Distributions of $M_{p\bar{p}}-2m_{p}$ and fit result
   corresponding to the upper limit on the branching fraction at the 90\% C.L..
  The dots with error
bars represent the data, the black solid line is the global
fit result, the red dashed-dotted line is the $X(p\bar p)$ signal, and the
blue long-dashed-dotted line denotes the non-resonant background.
      }
\end{figure}

Figure~\ref{Mppbar} depicts the $p\bar{p}$ invariant-mass distribution
for the events with a $\kk$ invariant mass that falls within the $\phi$ mass region
   (1.005~GeV/$c^{2} < M_{\kk} <$ 1.035~GeV/$c^{2}$),
where no evident enhancement near the $p\bar{p}$-mass threshold is visible.
It is found that the events from the phase-space process together with other
background components provide a good description of the data, which is shown in Fig.~\ref{Mppbar}.
Therefore, an upper limit for the $X(\ppbar)$ production rate can be measured.
For $\jpsi\to\ppphi$, we divide $\ppbar$ invariant-mass spectrum into 9 bins in the region of
[1.876, 2.056] GeV/$c^{2}$. With the same procedure as described above, the number of the $\phi$
events in each bin can be obtained by fitting to the corresponding $\kk$-mass spectrum.
Subsequently, the non-$\phi$-background-subtracted $M_{\ppbar}$ distribution is obtained as shown in Fig.~\ref{Fit},
 where the errors are statistical only,
 and $m_p$ is the nominal mass of proton~\cite{pdg}.

The spin ($J$) and parity ($P$) of $X(p\bar{p})$ have been determined by an
amplitude analysis of $J/\psi \to \gamma p\bar{p}$ decay and resulted in $J^{PC} = 0^{-+}$~\cite{cite04}.
In our analysis, we parametrize the $X(p\bar{p})$ signal by an efficiency-weighted $S$-wave BW function,

\begin{equation}
BW(M) \simeq \frac{f_{FSI} \times q^{2L+1}\kappa^3}{(M^2-M^2_0)^2 + M^2_0\Gamma^2_0} \times \varepsilon_{rec}(M) ,
\end{equation}

\noindent where $M$ is the $p\bar{p}$ invariant mass, the parameter $f_{FSI}$ accounts for the effect of the FSI,
$q$ is the momentum of the proton in the $p\bar{p}$ rest frame,
$\kappa$ is the momentum of $\phi$ in the $\psi(3686)$ rest frame,
$L = 0$ is the relative orbital angular momentum of the $p\bar{p}$ system,
$M_0$ and $\Gamma_0$ are the mass and width of $X(p\bar{p})$, respectively, 
which are fixed to those in Ref.~\cite{cite04}.  
$\varepsilon_{rec}(M)$ is the mass-dependent detection efficiency which is obtained from MC simulations of
$\psi(3686) \to X(p\bar{p})\phi \to p\bar{p}\phi$.
We ignore possible interference effects of the $X(p\bar{p})$ resonance with non-resonant background contributions.

To determine the upper limit on the size of the $p\bar p $ enhancement,
a series of binned least-$\chi^2$ fits are performed to the background-subtracted
$p\bar{p}$-mass spectrum with the expected signal. Fit-related uncertainties are
included by considering the following three aspects:
(a) the $X(p\bar{p})$ signal is described by excluding the FSI factor with $f_{FSI}=1$ or taking into account
the J\"ulich FSI value as described in Ref.~\cite{cite15};
(b) the non-resonant background is represented by the shape obtained from
the $\psi(3686) \to p\bar{p}\phi$ MC simulation or parameterized by a function
of $f(\delta) = N(\delta^{1/2} + a_1\delta^{3/2} + a_2\delta^{5/2})$ ($\delta = M_{p\bar{p}} - 2m_{p}$,
$a_1$ and $a_2$ are free parameters); and
(c) the fit is performed in the range of [0.00, 0.18] GeV/$c^{2}$ or [0.00, 0.20] GeV/$c^{2}$.
Therefore, there are eight alternative fit scenarios.
In the variations, the fit taking into account the FSI,
with the background parameterized by the function of $f(\delta)$ in the range [0.0, 0.18] GeV/$c^{2}$,
gives the maximum number of $X(p\bar{p})$ candidates, 20.6, at the 90\% C.L..
The corresponding fitting plot is shown in Fig.~\ref{Fit},
 and the upper limit on the branching fraction is determined by,
\begin{equation}
\begin{split}
\mathcal{B}(\psi(3686) &\to X(p\bar{p})\phi \to p\bar{p}\phi)\\
&< \frac{N^{UL}}{N_{\psi(3686)} \times \mathcal{B}(\phi \to K^+K^-) \times \varepsilon},
\end{split}
\end{equation}

\noindent where $N^{UL}$ is the maximum number of $X(p\bar{p})$ events.
To be conservative,
the multiplicative uncertainties listed in Table~\ref{sys} are considered by convoluting the
 normalised $\chi^{2}$ distribution with a Gaussian function.
 The detection efficiency, $\varepsilon$, is obtained from MC simulations,
 and is determined to be 58.9\%.
The upper limit on the branching fraction of $\psi(3686) \to X(p\bar{p})\phi \to p\bar{p}\phi$ at the 90\%
C.L. is calculated to be $1.82 \times 10^{-7}$.

\section{SUMMARY}
Using a sample of $4.48 \times 10^8$ $\psi(3686)$ events accumulated with the BESIII detector,
we present a study of the decay $\psi(3686) \to p\bar{p}\phi$. The  branching fraction
 of $\psi(3686) \to p\bar{p}\phi$ is measured for the first time and
 it is found to be $[6.06 \pm 0.38($stat.$) \pm 0.48($syst.$)] \times 10^{-6}$. 
 With the previously published branching-fraction measurement of $J/\psi\to p \bar{p}\phi$~\cite{cite20},
 the ratio $Q=\frac{\mathcal{B}(\psi(3686)\to p\bar{p}\phi)}{\mathcal{B}(J/\psi\to p\bar{p}\phi)}$ is determined to be $(11.6\pm0.7\pm1.2)\%$. 
  With the same approach as given in Ref.~\cite{ppeta}, we also present the ratio by taking the phase spaces of
 $J/\psi/\psi(3686)\to p \bar{p}\phi$ into account.
 The phase-space ratio of them is determined to be $\Omega_{\psi(3686)\to p\bar{p}\phi}/\Omega_{J/\psi\to p\bar{p}\phi}=11.9$.
 By taking this into consideration, the $Q$ value becomes $(0.97\pm0.06\pm0.10)\%$, which indicates that the `$12\%$ rule' is violated significantly.
No evidence for an enhancement near the $p\bar{p}$-mass threshold is found and
the upper limit on the branching fraction of
$\psi(3686) \to X(p\bar{p})\phi \to p\bar{p}\phi$ is determined to be
 $1.82 \times 10^{-7}$ at the 90\% C.L..\\
\vspace{4pt}

\begin{acknowledgments}

The BESIII collaboration thanks the staff of BEPCII and the IHEP computing center for their strong support. This work is supported in part by National Key Basic Research Program of China under Contract No. 2015CB856700; National Natural Science Foundation of China (NSFC) under Contracts Nos. 11335008, 11425524, 11625523, 11635010, 11735014, 11565006; the Chinese Academy of Sciences (CAS) Large-Scale Scientific Facility Program; the CAS Center for Excellence in Particle Physics (CCEPP); Joint Large-Scale Scientific Facility Funds of the NSFC and CAS under Contracts Nos. U1332201, U1532257, U1532258, U1732263; CAS Key Research Program of Frontier Sciences under Contracts Nos. QYZDJ-SSW-SLH003, QYZDJ-SSW-SLH040; 100 Talents Program of CAS; INPAC and Shanghai Key Laboratory for Particle Physics and Cosmology; German Research Foundation DFG under Contract No. Collaborative Research Center CRC 1044; Istituto Nazionale di Fisica Nucleare, Italy; Koninklijke Nederlandse Akademie van Wetenschappen (KNAW) under Contract No. 530-4CDP03; Ministry of Development of Turkey under Contract No. DPT2006K-120470; National Science and Technology fund; The Swedish Research Council; U. S. Department of Energy under Contracts Nos. DE-FG02-05ER41374, DE-SC-0010118, DE-SC-0012069; University of Groningen (RuG) and the Helmholtzzentrum fuer Schwerionenforschung GmbH (GSI), Darmstadt\end{acknowledgments}


\begin{thebibliography}{99}



\bibitem{cite01} J. Z. Bai {\it et al.} (BES Collaboration), Phys. Rev. Lett. {\bf 91}, 022001 (2003).
\bibitem{cite02} J. P. Alexander {\it et al.} (CLEO Collaboration), Phys. Rev. D {\bf 82}, 092002 (2010).
\bibitem{cite03} M. Ablikim {\it et al.} (BESIII Collaboration), Chin. Phys. C {\bf 34}, 421 (2010).
\bibitem{cite04} M. Ablikim {\it et al.} (BESIII Collaboration), Phys. Rev. Lett. {\bf 108}, 112003 (2012).
\bibitem{cite05} S. B. Athar {\it et al.} (CLEO Collaboration), Phys. Rev. D {\bf 73}, 032001 (2006).
\bibitem{cite06} M. Ablikim {\it et al.} (BES Collaboration), Eur. Phys. J. C {\bf 53}, 15 (2008).
\bibitem{cite09} M. Ablikim {\it et al.} (BESIII Collaboration), Phys. Rev. Lett. {\bf 106}, 072002 (2011).
\bibitem{cite10} M. Ablikim {\it et al.} (BESII Collaboration), Phys. Rev. Lett. {\bf 95}, 262001 (2005).
\bibitem{cite11} A. Datta and P. J. O'Donnell, Phys. Lett. B {\bf 567}, 273 (2003).
\bibitem{cite12} M. L. Yan, S. Li, B. Wu and B. Q. Ma, Phys. Rev. D {\bf 72}, 034027 (2005).
\bibitem{cite13} M. Abud, F. Buccella, and F. Tramontano, Phys. Rev. D {\bf 81}, 074018 (2010).
\bibitem{cite14} B. S. Zou and H. C. Chiang, Phys. Rev. D {\bf 69}, 034004 (2004).
\bibitem{cite15} A. Sibirtsev, J. Haidenbauer, S. Krewald, U. Meissner, and A. W. Thomas, Phys. Rev. D {\bf 71}, 054010 (2005).
\bibitem{cite20} M. Ablikim {\it et al.} (BESIII Collaboration), Phys. Rev. D {\bf 93}, 052010 (2016).
\bibitem{citeBESI} M. Ablikim {\it et al.} (BESIII Collaboration), Phys. Rev. D {\bf 67}, 052002 (2003).
\bibitem{cite21} R. A. Briere {\it et al.} (CLEO Collaboration), Phys. Rev. Lett. {\bf 95}, 062001 (2005).
\bibitem{cite22} M. Ablikim {\it et al.} (BESIII Collaboration), Chin. Phys. C {\bf 42}, 023001 (2018).
\bibitem{cite23} T. Appelquist and H. D. Politzer, Phys. Rev. Lett. {\bf 34}, 43 (1975).
\bibitem{Ablikim:2009aa}
  M.~Ablikim {\it et al.} (BESIII Collaboration),
  Nucl. Instrum.
Methods Phys. Res., Sect. A {\bf 614}, 345 (2010).

\bibitem{Yu:IPAC2016-TUYA01}
   C.~H.~Yu {\it et al.},
  Proceedings of IPAC2016, Busan, Korea, 2016,
  doi:10.18429/JACoW-IPAC2016-TUYA01.

\bibitem{geant4}
  S.~Agostinelli {\it et al.} (GEANT4 Collaboration),
   Nucl. Instrum.
Methods Phys. Res., Sect. A {\bf 506}, 250 (2003).

\bibitem{ref:kkmc}
  S.~Jadach, B.~F.~L.~Ward and Z.~Was,
    Phys.\ Rev.\ D {\bf 63}, 113009 (2001);
   Comput.\ Phys.\ Commun.\  {\bf 130}, 260 (2000).

\bibitem{ref:evtgen}
  D.~J.~Lange,
  Nucl. Instrum. Methods Phys. Res., Sect. A {\bf 462}, 152 (2001);
  R.~G.~Ping,
  Chin. Phys. C {\bf 32}, 599 (2008).

\bibitem{pdg}
M. Tanabashi {\it et al.} (Particle Data Group), Phys. Rev. D {\bf98}, 030001 (2018).

\bibitem{ref:lundcharm}
  J.~C.~Chen, G.~S.~Huang, X.~R.~Qi, D.~H.~Zhang and Y.~S.~Zhu,
   Phys.\ Rev.\ D {\bf 62}, 034003 (2000);
  R.~L.~Yang, R.~G.~Ping and H.~Chen,
   Chin.\ Phys.\ Lett.\  {\bf 31}, 061301 (2014).

\bibitem{photos}
  E.~Richter-Was,
   Phys.\ Lett.\ B {\bf 303}, 163 (1993).

\bibitem{ref:boss} W.~D.~Li, H.~M.~Liu {\it et al.}, in proceeding of CHEP06, Mumbai, India, 2006 edited by Sunanda Banerjee (Tata Institute of Fundamental Reserach, Mumbai, (2006).


\bibitem{N3773} M. Ablikim {\it et al.} (BESIII Collaboration), Phys. Lett. B {\bf 753}, 629 (2016).


\bibitem{cite32} H. Albrecht {\it et al.} (ARGUS Collaboration), Phys. Lett. B {\bf 340}, 217 (1994).
\bibitem{ywccite33} M. Ablikim {\it et al.} (BESIII Collaboration), Phys. Rev. D {\bf83}, 112005 (2011).
\bibitem{ywccite34} M. Ablikim {\it et al.} (BESIII Collaboration), Phys. Rev. D {\bf85}, 092012 (2012).

\bibitem{ppeta}
  M.~Ablikim {\it et al.} [BESIII Collaboration],
  Phys.\ Rev.\ D {\bf 99}, 032006 (2019).

\end{thebibliography}
\end{document}